\documentclass[12pt]{article}

\usepackage[margin=1in]{geometry}
\usepackage{setspace}
\usepackage{amsmath,amssymb,amsthm,mathtools}
\usepackage{booktabs,tabularx,array,threeparttable}
\usepackage{graphicx}
\usepackage{natbib}
\usepackage{xurl}
\usepackage{hyperref}
\hypersetup{hidelinks}
\usepackage{enumitem}
\usepackage{authblk}

\newtheorem{theorem}{Theorem}
\newtheorem{proposition}{Proposition}
\newtheorem{corollary}{Corollary}
\newtheorem{lemma}{Lemma}
\newtheorem{assumption}{Assumption}
\newtheorem{remark}{Remark}

\newcommand{\Pbb}{\mathbb{P}}

\begin{document}

\title{Finite-Boundary Reduction and Exact Verification of Strong Familywise Error in Active-Count-Coupled Multi-Arm Efficacy--Toxicity Monitoring}
\author[1,2]{Masahiro Kojima}
\author[2,3]{Hisato Sunami}
\author[4]{Kentaro Takeda}
\affil[1]{Department of Data Science for Business Innovation, Faculty of Science and Engineering, Chuo University, Tokyo, Japan}
\affil[2]{Biometrics Department, Research Division, Kyowa Kirin Co., Ltd., Tokyo, Japan}
\affil[3]{Department of Biomedical Statistics, Graduate School of Medicine, The University of Osaka, Osaka, Japan}
\affil[4]{Astellas Pharma Global Development Inc., Northbrook, IL, USA.}
\date{}
\singlespacing
\maketitle

\noindent\textbf{Running head:} Strong error verification under active-count coupling

\noindent\textbf{Correspondence:} Masahiro Kojima, Department of Data Science for Business Innovation, Faculty of Science and Engineering, Chuo University, 1-13-27 Kasuga, Bunkyo-ku, Tokyo 112-8551, Japan. Email: mkojima263@g.chuo-u.ac.jp

\noindent\textbf{Funding:} This work was supported by the Japan Society for the Promotion of Science KAKENHI [JP26K21185 to M.K.].

\noindent\textbf{Conflict of interest:} Masahiro Kojima and Hisato Sunami are employees of Kyowa Kirin Co., Ltd. Kentaro Takeda is an employee of Astellas Pharma Global Development Inc. The authors declare no other conflicts of interest.

\noindent\textbf{Acknowledgements:} The authors thank their colleagues for helpful discussions. OpenAI's ChatGPT was used only to assist with manuscript proofreading and as an additional check of the simulation code and numerical results. The authors independently reviewed and verified the manuscript, code, numerical results, references, and scientific conclusions and take full responsibility for the work.

\doublespacing
\begin{abstract}
Randomized dose-optimization trials may screen several candidate doses using binary efficacy and toxicity outcomes. A dose is inadmissible if efficacy is insufficient or toxicity is excessive, so each dose-specific null hypothesis is a union region and strong familywise error control must hold across arbitrary mixtures of inadmissible and promising doses. We study exact verification of multistage monitoring rules in which armwise decisions may be coupled through the number of active arms remaining at each interim analysis. For uncoupled rules, we derive an exact product representation and show that a complete-null boundary configuration is least favourable. Under active-count coupling, this factorization no longer holds because a promising dose can remain active and alter future boundaries for null doses. Assuming independent sampling across arms, prespecified per-dose analysis schedules, analysis timing not driven by the observed efficacy or toxicity outcomes, arm-specific monitoring statistics, and stagewise monotonicity, we establish an exact finite-boundary characterization without parametric restrictions on the within-arm joint efficacy–toxicity distribution. Each dose need only be evaluated at an efficacy-null boundary, a toxicity-null boundary, or a maximally favourable alternative. A deterministic finite-state recursion then verifies a fixed decision table without Monte Carlo error. Numerical studies confirmed the prespecified strong familywise error control and showed that active-count coupling can shift the least-favourable configuration away from the complete null while improving joint retention of multiple promising doses. A published randomized dose-ranging trial was used as a clinical illustration of how the framework could be prospectively implemented. The framework separates monitoring-rule construction from rigorous error verification.
\end{abstract}


\section{Introduction}\label{sec:introduction}

Contemporary oncology dose optimization increasingly emphasizes prospective comparison of multiple candidate dosages with consideration of both efficacy and safety. The U.S. Food and Drug Administration's Project Optimus and subsequent dosage-optimization guidance encourage characterization of the benefit--risk profile of candidate dosages before marketing application \citep{FDAProjectOptimus2021,gao2024realizing,FDADoseOptimization2024}. Although this regulatory framework also considers tolerability, pharmacokinetics, pharmacodynamics, patient-reported outcomes, and other clinical information, the present work focuses on binary efficacy and toxicity endpoints. Multi-arm multistage phase II trials are attractive in this setting because clearly unpromising doses can be dropped while other doses continue. Their use raises a multiplicity question that is distinct from the Bayesian or frequentist form of the armwise monitoring statistic: how should the probability of retaining one or more truly inadmissible doses be controlled over the full data-generating parameter space?

A candidate dose is inadmissible if its efficacy probability does not exceed a prespecified minimum acceptable level or its toxicity probability reaches or exceeds a prespecified maximum acceptable level. The dose-level null is therefore a continuous union region rather than a single point. At the trial level, strong familywise error rate (FWER) control requires protection for every possible subset of true null doses, including configurations in which inadmissible and genuinely promising doses coexist. This distinction becomes important when the decision for one dose depends on the status of other doses. In an active-count-dependent design, a promising dose can remain active, keep the active count high, and thereby alter later monitoring boundaries applied to a null dose. Consequently, a configuration in which all doses are null need not be least favourable: a genuinely promising dose can contribute indirectly to the strong-FWER problem by changing the monitoring boundary faced by a null dose.

Existing methods address different components of this problem. BOP2-based approaches provide multistage efficacy--toxicity monitoring, and the multi-arm extension of \citet{mulier2024bayesian} allows monitoring thresholds to depend on the evolving number of active arms, but its calibration is based on prespecified reference configurations rather than a general strong-FWER characterization over the full efficacy--toxicity union null. BOP2-TE provides exact treatment-level error calculations at several clinically meaningful point null configurations \citep{chen2026bop2}, whereas MERIT calibrates randomized dose-optimization decisions over a finite collection of ordered efficacy--toxicity configurations \citep{yang2024design}. More recently, \citet{tabata2026dose} established dose-specific error control throughout the continuous efficacy--toxicity union null using a fixed-sample construction that can be combined across doses to obtain strong FWER control. A detailed comparison of the inferential scope and adaptive structure of these approaches is provided in Supplementary Table~S1. To our knowledge, these approaches do not simultaneously provide an exact strong-FWER characterization over the full continuous efficacy--toxicity union null for arbitrary mixtures of inadmissible and promising doses, repeated decision paths coupled through the evolving active set, and an unrestricted within-patient joint efficacy--toxicity distribution.

To address this gap, we develop a general exact verification framework for a broad class of multistage monitoring rules, including rules coupled through the evolving active count. For uncoupled rules, in which each dose decision depends only on that dose's own outcome history, we derive an exact product representation of the strong FWER and show that its worst case is attained at a complete-null boundary configuration. For rules coupled through the evolving number of active arms, we show, under independent arm sampling, prespecified per-dose analysis schedules, analysis timing not driven by the observed efficacy or toxicity outcomes, arm-specific monitoring statistics, and stagewise monotonicity, that the strong-FWER supremum over unrestricted valid arm-specific four-category efficacy--toxicity distributions reduces exactly to a finite set of boundary configurations. Each dose need only be represented by an efficacy-null boundary state, a toxicity-null boundary state, or a maximally favourable alternative state. Once a decision table is fixed, deterministic finite-state recursion can evaluate all required configurations without Monte Carlo error. Numerical studies are used to evaluate exact error control and operating characteristics for illustrative uncoupled and active-count-coupled rules, and a published randomized dose-ranging trial provides a clinical illustration of how the framework could be prospectively specified in practice.

The remainder of the paper is organized as follows. Section~\ref{sec:methods} defines the monitoring framework and strong FWER criterion and establishes the uncoupled product representation and the finite-boundary reductions. Section~\ref{sec:numerical-evaluation} describes the illustrative monitoring rules, their calibration and exact verification, and the resulting operating characteristics. Section~\ref{sec:abroad-case} gives the prospective clinical illustration based on a published randomized dose-ranging trial. Section~\ref{sec:discussion} discusses the interpretation, scope, limitations, and implications of the proposed verification framework.

\section{Methods}\label{sec:methods}

We first describe the multi-arm efficacy--toxicity monitoring framework and define the strong familywise error criterion. We then state the conditions that yield pathwise monotonicity and establish the exact product representation for uncoupled armwise rules. We next show how active-count coupling changes the least-favourable-configuration problem, leading to a finite-boundary reduction over the unrestricted parameter space and a further reduction under nondecreasing toxicity. Proofs of the technical results and computational details are provided in the Supplementary Material.

\subsection{Trial framework and monitoring rules}\label{sec:trial-framework}

We consider $K\geq2$ experimental arms evaluated with binary efficacy and toxicity endpoints. For patient $i$ on arm $k$, let $Y_{E,ik}\in\{0,1\}$ and $Y_{T,ik}\in\{0,1\}$ denote efficacy and toxicity, respectively. We use the standard four-category order $(0,0)$, $(0,1)$, $(1,0)$, and $(1,1)$ and write the arm-specific probability vector as $\boldsymbol\pi_k=(\pi_{k00},\pi_{k01},\pi_{k10},\pi_{k11})^{\mathsf T}$. The marginal probabilities are $p_{E,k}=\pi_{k10}+\pi_{k11}$ and $p_{T,k}=\pi_{k01}+\pi_{k11}$, where $\pi_{ket}=\Pbb(Y_{E,ik}=e,Y_{T,ik}=t), e,t\in\{0,1\}$. Conditional on the arm-specific probability vectors, patient outcomes are independent and identically distributed within each arm and independent across arms; efficacy and toxicity may be associated arbitrarily within a patient. Independent arm-specific Dirichlet priors are used, with no cross-arm borrowing. The posterior marginal distributions of $p_{E,k}$ and $p_{T,k}$ are beta distributions obtained by Dirichlet aggregation.

Let $\phi_E\in(0,1)$ be the minimum acceptable efficacy probability and $\phi_T\in(0,1)$ the maximum acceptable toxicity probability. The dose-level null and alternative are
\begin{align*}
H_{0k}&:\quad p_{E,k}\leq\phi_E\ \text{or}\ p_{T,k}\geq\phi_T,\\
H_{1k}&:\quad p_{E,k}>\phi_E\ \text{and}\ p_{T,k}<\phi_T.
\end{align*}
For ordered doses we may additionally restrict the parameter space by $p_{T,1}\leq\cdots\leq p_{T,K}$, but the primary finite-reduction theorem below does not require this restriction.

We consider a $J$-stage per-arm design with fixed analysis sample sizes $0<n_1<\cdots<n_J=N$. Each active arm is analyzed when its cumulative evaluable sample size reaches one of these prespecified values. Analyses $j=1,\ldots,J-1$ are interim analyses, and analysis $J$ is the final analysis. Different arms need not reach these analysis sizes at the same calendar time. The accrual and analysis-timing process that determines when the prespecified sample sizes are reached is assumed not to depend on the efficacy or toxicity outcomes. Conditional on a realization of this process, let $s=1,\ldots,S$ index the resulting potential analysis opportunities across the trial, with analyses occurring at the same time treated as a common opportunity. If an arm has already been dropped, its later potential analysis opportunities are skipped. Let $\mathcal A_s^{-}$ denote the set of arms under monitoring immediately before opportunity $s$ and let $m_s=|\mathcal A_s^{-}|$. If arm $k$ undergoes its $j$th analysis at opportunity $s$, define posterior evidence for insufficient efficacy and excessive toxicity, based on the data $\mathcal D_k(n_j)$ accumulated on arm $k$ through its $j$th analysis, by
\begin{align*}
F_{E,kj}&=\Pbb\{p_{E,k}\leq\phi_E\mid\mathcal D_k(n_j)\},\\
F_{T,kj}&=\Pbb\{p_{T,k}\geq\phi_T\mid\mathcal D_k(n_j)\}.
\end{align*}
A convenient implementation uses an endpoint-specific version of the active-count-dependent threshold proposed by \citet{mulier2024bayesian},
\begin{equation}
C_r(n,m)=1-\frac{K+1-m-\lambda_r}{K+1-m}\left(\frac{n}{N}\right)^{\gamma_r},
\qquad r\in\{E,T\},
\label{eq:monitoring-threshold}
\end{equation}
where $0<\lambda_r<1$ and $\gamma_r>0$, and drops arm $k$ at opportunity $s$ when $F_{E,kj}>C_E(n_j,m_s)$ or $F_{T,kj}>C_T(n_j,m_s)$. The theory does not require this parametric form. Every fixed posterior rule induces integer boundaries: let $e_j(m)$ be the largest efficacy count that triggers futility and $t_j(m)$ the smallest toxicity count that triggers excessive-toxicity stopping. The retain region is
\begin{equation}
x_{E,k}(n_j)\geq e_j(m_s)+1,\qquad
x_{T,k}(n_j)\leq t_j(m_s)-1.
\label{eq:retain-region}
\end{equation}
A directly specified integer rule is also allowed if, for every analysis,
\begin{equation}
e_j(1)\geq\cdots\geq e_j(K),\qquad
t_j(1)\leq\cdots\leq t_j(K).
\label{eq:integer-monotonicity}
\end{equation}
Thus, increasing the number of active arms cannot make another arm easier to stop. The complete cutoff or integer table is fixed prospectively; the observed active count only selects a prespecified row.

At its final analysis an arm is declared promising if it remains under monitoring and satisfies the retain region. Let $R_k$ denote this event. After its final analysis, the arm no longer contributes to subsequent active counts.

\subsection{Strong familywise error and pathwise monotonicity}\label{sec:strong-fwer}

For a true configuration $\boldsymbol\Theta=(\boldsymbol\pi_1,\ldots,\boldsymbol\pi_K)$, define the set of truly inadmissible arms by
\[
\mathcal I_0(\boldsymbol\Theta)
=
\left\{
k:
p_{E,k}\leq\phi_E
\ \text{or}\
p_{T,k}\geq\phi_T
\right\}.
\]
For the dose-screening decision, we define the familywise error rate as
\begin{equation*}
\operatorname{FWER}(\boldsymbol\Theta)
=
\Pbb_{\boldsymbol\Theta}
\left(
\bigcup_{k\in\mathcal I_0(\boldsymbol\Theta)}R_k
\right),
\end{equation*}
where the union over an empty index set is interpreted as the empty event, so that $\operatorname{FWER}(\boldsymbol\Theta)=0$ when $\mathcal I_0(\boldsymbol\Theta)=\varnothing$. Strong control over a prespecified parameter space $\mathfrak T$ requires
\[
\sup_{\boldsymbol\Theta\in\mathfrak T}
\operatorname{FWER}(\boldsymbol\Theta)
\leq\alpha.
\]

The complete-null region within $\mathfrak T$ is
$\mathfrak T_0=\left\{\boldsymbol\Theta\in\mathfrak T:
\mathcal I_0(\boldsymbol\Theta)=\{1,\ldots,K\}\right\}$.
Because this region is itself composite, we use \emph{composite weak control} to mean
\[
\sup_{\boldsymbol\Theta\in\mathfrak T_0}
\operatorname{FWER}(\boldsymbol\Theta)
\leq\alpha,
\]
whereas control at a single prespecified configuration
$\boldsymbol\Theta_0\in\mathfrak T_0$ is only pointwise control.

Let
\[
\Delta_3
=
\left\{
\boldsymbol\pi\in[0,1]^4:
\sum_{\ell=1}^{4}\pi_\ell=1
\right\}
\]
denote the three-dimensional probability simplex for the four-category outcome. The unrestricted parameter space is
\begin{equation*}
\mathfrak T_U
=
\Delta_3^K
=
\left\{
(\boldsymbol\pi_1,\ldots,\boldsymbol\pi_K):
\boldsymbol\pi_k\in\Delta_3,\ k=1,\ldots,K
\right\}.
\end{equation*}
Thus, within each arm, the efficacy probability, toxicity probability, and their within-patient association are unrestricted subject only to the validity of the four-category probability vector. The across-patient and across-arm independence assumptions stated in Section~\ref{sec:trial-framework} concern the sampling law rather than the definition of $\mathfrak T_U$. Although no toxicity ordering is required for the unrestricted result, a nondecreasing dose--toxicity relationship, including possible plateaus, is a common working assumption in oncology dose-finding \citep{lin2023bayesian,GuoYuan2023}. Accordingly, when such an ordering is scientifically justified for the candidate doses, define
\begin{equation*}
\mathfrak T_M
=
\left\{
\boldsymbol\Theta\in\mathfrak T_U:
p_{T,1}\leq\cdots\leq p_{T,K}
\right\}.
\end{equation*}

For each per-arm analysis index $j$ and active-arm count $m\in\{1,\ldots,K\}$, let $\mathcal R_j(m)$ denote the set of efficacy--toxicity count pairs for which an active arm passes its $j$th analysis when $m$ arms are under monitoring immediately before the corresponding analysis opportunity. At an interim analysis ($j<J$), passing the rule means that the arm remains under monitoring; at the final analysis ($j=J$), passing means that the arm is declared promising.

For readability, a possible arm index is suppressed in $\mathcal R_j(m)$ and in the corresponding integer boundaries. Prospectively fixed arm-specific pass regions are permitted provided that the conditions below hold armwise. The exchangeability reduction later requires these design features to be invariant under relabelling of the arms.

\begin{assumption}\label{ass:monotone-conduct}
The design satisfies the following conditions:
\begin{enumerate}[label=(\roman*),leftmargin=2.5em]
\item All $K$ experimental arms are active before the first analysis, no new experimental arm is added thereafter, and each active arm follows the fixed per-arm analysis schedule $n_1<\cdots<n_J=N$ unless it is dropped by its own monitoring rule.

\item Both binary endpoints required at an analysis are observed for the evaluable patients used in that decision.

\item The monitoring statistic for an arm is based only on cumulative marginal counts \((X_{E,kj},X_{T,kj})\) from that arm, and information from other arms may enter its decision only through the number of active arms, without cross-arm posterior borrowing, pooling, or isotonic modification of the monitoring statistic.

\item Each active arm is analyzed when it reaches a prespecified per-arm analysis sample size. The accrual and analysis-timing process that determines when these sample sizes are reached does not depend on the efficacy or toxicity outcomes. Later potential analysis opportunities for an arm that has already been dropped are skipped. If multiple arms are analyzed at the same opportunity, all corresponding decisions use the active-arm count immediately before any of those decisions are applied. Arms are not removed for early efficacy success before the final analysis.

\item For every per-arm analysis index $j$, the pass region $\mathcal R_j(m)$ satisfies the following two monotonicity conditions:
\begin{align}
(x_E,x_T)\in\mathcal R_j(m),\quad
x'_E\geq x_E,\quad
x'_T\leq x_T
&\quad\Longrightarrow\quad
(x'_E,x'_T)\in\mathcal R_j(m),
\label{eq:outcome-monotonicity}\\
(x_E,x_T)\in\mathcal R_j(m),\quad
m'\geq m
&\quad\Longrightarrow\quad
(x_E,x_T)\in\mathcal R_j(m').
\label{eq:active-count-monotonicity}
\end{align}
\end{enumerate}
\end{assumption}

The two monotonicity conditions in Assumption~\ref{ass:monotone-conduct}(v) serve complementary roles. Equation~\eqref{eq:outcome-monotonicity} ensures that, for a fixed number of active arms, increasing the cumulative efficacy count or decreasing the cumulative toxicity count cannot change a passing decision into a failing decision. Equation~\eqref{eq:active-count-monotonicity} ensures that, for fixed arm-specific data, the same reversal cannot occur when more arms remain active. Together with conditions (i)--(iv), these conditions yield pathwise active-set inclusion across successive analysis opportunities, as established in Lemma~\ref{lem:active-set-inclusion} and subsequently used in the finite-boundary reduction.

For the integer-boundary representation, the stage-$j$ pass region is given by \eqref{eq:retain-region}. Hence \eqref{eq:outcome-monotonicity} follows directly from the count-based decision rule, and \eqref{eq:active-count-monotonicity} follows from the boundary ordering in \eqref{eq:integer-monotonicity}.

For the posterior-probability representation defined in Section~\ref{sec:trial-framework}, the beta marginal posteriors are monotone in the corresponding cumulative counts at fixed $n_j$:
\begin{align*}
x'_E\geq x_E
&\quad\Longrightarrow\quad
F_{E,kj}(x'_E)\leq F_{E,kj}(x_E),\\
x'_T\leq x_T
&\quad\Longrightarrow\quad
F_{T,kj}(x'_T)\leq F_{T,kj}(x_T).
\end{align*}
Thus \eqref{eq:outcome-monotonicity} holds for the posterior-probability rule. In addition, \eqref{eq:active-count-monotonicity} holds provided that
\begin{equation*}
m'\geq m
\quad\Longrightarrow\quad
C_r(n_j,m')\geq C_r(n_j,m),
\qquad
r\in\{E,T\}.
\end{equation*}
For the active-count threshold in Equation~\eqref{eq:monitoring-threshold}, this condition holds because $C_r(n,m)$ is nondecreasing in $m$ whenever $0<\lambda_r<1$.

\begin{lemma}\label{lem:active-set-inclusion}
Consider two complete outcome arrays under the same monitoring design and the same realization of the accrual and analysis-timing process:
\[
\left\{
(Y_{E,ik},Y_{T,ik}):
i=1,\ldots,N,\ k=1,\ldots,K
\right\},
\]
and
\[
\left\{
(\tilde Y_{E,ik},\tilde Y_{T,ik}):
i=1,\ldots,N,\ k=1,\ldots,K
\right\}.
\]
Let $\mathcal A_s^{-}$ and $\tilde{\mathcal A}_s^{-}$ denote the sets of arms under monitoring immediately before analysis opportunity $s$ under the original and favourable outcome paths, respectively, and let $\mathcal A_s^{+}$ and $\tilde{\mathcal A}_s^{+}$ denote the corresponding sets immediately after all decisions at that opportunity have been applied. If
\[
\tilde Y_{E,ik}\geq Y_{E,ik},
\qquad
\tilde Y_{T,ik}\leq Y_{T,ik}
\]
for every patient index $i$ and arm $k$, then under Assumption~\ref{ass:monotone-conduct},
\[
\mathcal A_s^{-}\subseteq\tilde{\mathcal A}_s^{-},
\qquad
\mathcal A_s^{+}\subseteq\tilde{\mathcal A}_s^{+},
\qquad
s=1,\ldots,S.
\]
In particular, if $R_k$ occurs under the original outcome path, then the corresponding event $\tilde R_k$ occurs under the favourable outcome path.
\end{lemma}

A proof is provided in Supplementary Section~S1.1.

\subsection{Finite-boundary reduction}\label{sec:finite-reduction}

The preceding pathwise result provides the key step for reducing the strong-FWER maximization over the continuous parameter space. We next construct, for any configuration, a more favourable boundary configuration that preserves the set of inadmissible arms. The pathwise inclusion established above will then show that it is sufficient to maximize the false-selection probability over a finite set of such boundary configurations.

To state the reduction, define three extreme arm distributions
\begin{equation}
\begin{aligned}
L_E&=(1-\phi_E,0,\phi_E,0)^{\mathsf T},
&& (p_E,p_T)=(\phi_E,0),\\
L_T&=(0,0,1-\phi_T,\phi_T)^{\mathsf T},
&& (p_E,p_T)=(1,\phi_T),\\
A&=(0,0,1,0)^{\mathsf T},
&& (p_E,p_T)=(1,0).
\end{aligned}
\label{eq:boundary-states}
\end{equation}
The states $L_E$ and $L_T$ are null-boundary states that are maximally favourable in the other endpoint. The state $A$ is the maximally favourable alternative state: every patient has efficacy and no toxicity, so $(Y_E,Y_T)=(1,0)$ with probability one.

We call a monitoring rule \emph{uncoupled} if each arm is monitored independently of the other arms: whether arm $k$ is retained or ultimately declared promising is determined solely by its own efficacy--toxicity data and its prospectively fixed monitoring rule. Thus, the monitoring boundary and analysis schedule for arm $k$ do not change according to the outcomes or status of the other arms. In particular, the monitoring boundary does not depend on the number of active arms, and no shared-control information, cross-arm borrowing, pooling, or isotonic modification is used in the armwise decision. Formally, this means that the event $R_k$ depends only on the complete outcome sequence for arm $k$.

\begin{corollary}\label{cor:uncoupled-product}
Suppose that Assumption~\ref{ass:monotone-conduct} holds, the arm outcome paths are independent, and the monitoring rule is uncoupled. For arm $k$, define
\begin{equation*}
q_k^\star
=
\sup_{\substack{\boldsymbol\pi_k\in\Delta_3:\\
p_{E,k}\leq\phi_E\ \mathrm{or}\ p_{T,k}\geq\phi_T}}
\Pbb_{\boldsymbol\pi_k}(R_k).
\end{equation*}
Then
\begin{equation*}
q_k^\star
=
\max\left\{
\Pbb_{L_E}(R_k),
\Pbb_{L_T}(R_k)
\right\},
\end{equation*}
and
\begin{equation}
\sup_{\boldsymbol\Theta\in\mathfrak T_U}
\operatorname{FWER}(\boldsymbol\Theta)
=
1-\prod_{k=1}^{K}(1-q_k^\star).
\label{eq:uncoupled-product}
\end{equation}
The supremum is attained by a complete-null boundary configuration that assigns each arm to whichever of $L_E$ and $L_T$ attains $q_k^\star$. If the armwise rules are exchangeable, $q_k^\star=q^\star$ for all $k$ and Equation~\eqref{eq:uncoupled-product} becomes $1-(1-q^\star)^K$.
\end{corollary}

A proof is provided in Supplementary Section~S1.2.

\begin{remark}\label{rem:bonferroni-product-coupled}
For uncoupled independent arm paths, Equation~\eqref{eq:uncoupled-product} is the exact joint error probability. A uniform dose-level requirement such as $q_k^\star\leq\alpha/K$ yields the Bonferroni sufficient condition used by \citet{tabata2026dose}; that condition remains valid without arm independence. The exact product representation, however, relies on uncoupling. It also shows that introducing a truly promising arm cannot increase the false-retention probability of the remaining null arms and that a complete-null boundary configuration is least favourable.
\end{remark}

Active-count coupling changes this conclusion. Under an uncoupled rule, replacing a null arm by a genuinely promising arm cannot alter the monitoring path of the remaining null arms, which is why a complete-null boundary configuration is least favourable. Under active-count coupling, however, a promising arm can remain active, keep the active count larger, and enlarge the pass region for a null arm under Equation~\eqref{eq:active-count-monotonicity}. Thus, replacing a null arm by a favourable alternative can increase the false-retention probability of another null arm, and the factorization in Equation~\eqref{eq:uncoupled-product} is no longer available. The state $A$ defined in Equation~\eqref{eq:boundary-states} does not contribute directly to the false-retention event because it represents a truly promising arm. It is nevertheless required in the boundary set because such an arm can remain active at its interim analysis opportunities and thereby alter the later pass regions faced by null arms.

A boundary configuration for the $K$ arms is denoted by $\boldsymbol g=(g_1,\ldots,g_K)$, where $g_k$ is the distribution assigned to arm $k$. Define the finite boundary set
\begin{equation*}
\mathcal B_U
=
\left\{
\boldsymbol g=(g_1,\ldots,g_K):
g_k\in\{L_E,L_T,A\},\quad k=1,\ldots,K
\right\}
\setminus
\left\{
(A,\ldots,A)
\right\}.
\end{equation*}
For $\boldsymbol g\in\mathcal B_U$, the set of inadmissible arms is $\mathcal I_0(\boldsymbol g)=\left\{k:g_k\in\{L_E,L_T\}\right\}$, and $\Pbb_{\boldsymbol g}$ denotes probability under the $K$-arm configuration $\boldsymbol g$.

\begin{theorem}\label{thm:unrestricted-finite-reduction}
Under the sampling model in Section~\ref{sec:trial-framework} and Assumption~\ref{ass:monotone-conduct},
\begin{equation*}
\sup_{\boldsymbol\Theta\in\mathfrak T_U}
\operatorname{FWER}(\boldsymbol\Theta)
=
\max_{\boldsymbol g\in\mathcal B_U}
\Pbb_{\boldsymbol g}
\left(
\bigcup_{k\in\mathcal I_0(\boldsymbol g)}R_k
\right).
\end{equation*}
Hence the continuous strong-FWER problem reduces to $3^K-1$ labelled boundary configurations without imposing a dose-ordering assumption.
\end{theorem}

A proof is provided in Supplementary Section~S1.2.

\begin{remark}
The equality in Theorem~\ref{thm:unrestricted-finite-reduction} is an exact characterization, not a conservative upper bound. Corollary~\ref{cor:uncoupled-product} shows that the maximization collapses to an all-null product when arm paths are uncoupled. Under active-count coupling, mixed configurations containing the alternative state $A$ must remain in the finite set because a genuinely promising arm can alter the monitoring path followed by the null arms. Consequently, a complete-null configuration need not be least favourable.
\end{remark}

For ordered doses with nondecreasing toxicity, the arms satisfying $p_{T,k}<\phi_T$ form an initial segment. If the length of this segment is $q$, each arm in the segment is mapped to either $L_E$ or $A$, whereas each later arm is mapped to $L_T$. Let $\mathcal B_M$ denote the resulting set of boundary configurations after excluding $A^K$.

\begin{corollary}\label{cor:monotone-finite-reduction}
Under the sampling model in Section~\ref{sec:trial-framework}, Assumption~\ref{ass:monotone-conduct}, and nondecreasing toxicity,
\begin{equation*}
\sup_{\boldsymbol\Theta\in\mathfrak T_M}
\operatorname{FWER}(\boldsymbol\Theta)
=
\max_{g\in\mathcal B_M}
\Pbb_g
\left(
\bigcup_{k\in\mathcal I_0(g)}R_k
\right),
\end{equation*}
and
\[
|\mathcal B_M|=2^{K+1}-2.
\]
\end{corollary}

A proof is provided in Supplementary Section~S1.2.

\begin{proposition}\label{prop:exchangeability}
If priors, per-arm analysis schedules, decision rules, and the analysis-timing mechanism are exchangeable across arm labels, and information from other arms enters the monitoring rule only through the active count, then boundary configurations containing the same numbers of $L_E$, $L_T$, and $A$ arms have the same familywise error probability. The unrestricted boundary set therefore reduces to $K(K+3)/2$ non-all-$A$ equivalence classes. Under nondecreasing toxicity, the boundary set has the same number of equivalence classes when indexed by the length of the initial segment with toxicity below $\phi_T$ and the number of $L_E$ states within that segment.
\end{proposition}
A fixed asynchronous analysis order tied to particular arm labels need not satisfy this exchangeability condition, although the finite-boundary reduction in Theorem~\ref{thm:unrestricted-finite-reduction} remains valid.

A proof is provided in Supplementary Section~S1.2.

\section{Numerical evaluation}\label{sec:numerical-evaluation}
\subsection{Design settings and procedures}\label{sec:numerical-setting}
We conducted numerical studies to illustrate the exact verification framework and to compare the operating characteristics of representative uncoupled and active-count-coupled (AC-coupled) monitoring rules, with the latter using the endpoint-specific active-count-dependent threshold in Equation~\eqref{eq:monitoring-threshold}. 

The numerical evaluation considered $K=3$, $K=4$, and $K=5$ experimental doses. For every $K$, the maximum sample size was 60 evaluable patients per active dose, analyses occurred at 15, 30, 45, and 60 patients, decisions across active doses were simultaneous, the efficacy and toxicity reference values were $\phi_E=\phi_T=0.30$, and the trial-level strong-FWER constraint was 0.10. Each dose used a weak Dirichlet working prior with hyperparameters $(a_{00},a_{01},a_{10},a_{11})=(0.50,0.20,0.20,0.10)$ and total effective sample size one. These hyperparameters were chosen so that the induced marginal prior means for both efficacy and toxicity were 0.30, matching the corresponding reference values.

For each $K$, scenario $Q_q$, $q=1,\ldots,K$, assigned the first $q$ doses the common promising distribution $(p_E,p_T)=(0.60,0.15)$ with within-patient efficacy--toxicity odds ratio one, corresponding to the four-cell vector $(q_{00},q_{01},q_{10},q_{11})=(0.34,0.06,0.51,0.09)$. The remaining $K-q$ doses used the reference-null vector $(0.50,0.20,0.20,0.10)$. The value of $q$ is not assumed to be known by the design; it indexes alternative data-generating configurations used to evaluate operating characteristics. The monitoring rules were calibrated separately for each value of $K$ using the prespecified collection of $Q_q$ scenarios.

Candidate monitoring tables were selected by Monte Carlo search under the prespecified strong-FWER constraint. Within each value of $K$ and each rule family, common random numbers were used across candidate tables during search. The complete candidate-selection rule, including the prespecified fallback used when the primary retention target was not attainable, is given in the Supplementary Material. Once a table was selected, it was frozen and represented by its integer decision boundaries. Its strong FWER was then evaluated exactly by deterministic finite-state recursion over the finite boundary set established in Section~\ref{sec:finite-reduction}. Operating characteristics away from the boundary set were estimated using 100,000 simulated trials per scenario. Additional sensitivity analyses, with their scenario definitions fixed before execution, applied the same frozen tables without recalibration to alternative efficacy--toxicity margins, within-patient odds ratios of 0.5 and 2, and heterogeneous promising-dose profiles. These sensitivity analyses assess operating-characteristic robustness rather than re-establishing strong FWER, which is already covered over the full four-category parameter space by the exact verification result. Details of the search algorithm, deterministic recursion, independent audit, state-space scaling, and utility-based final recommendation are provided in the Supplementary Material.

\subsection{Exact strong-FWER verification}
\label{sec:exact-results}

Uncoupled and AC-coupled satisfied the prespecified strong-FWER constraint for every value of $K$. As guaranteed by Corollary~\ref{cor:uncoupled-product}, the Uncoupled maximum was attained under a complete-null boundary configuration: 0.0948 at $EEE$ for $K=3$, 0.0942 at $TTTT$ for $K=4$, and 0.0889 at $EEEEE$ for $K=5$. The AC-coupled maxima were 0.0934 at $TTT$ for $K=3$, 0.0969 at $TTTT$ for $K=4$, and 0.0956 at the mixed $ATTTT$ configuration for $K=5$. In the five-dose AC-coupled design, the largest complete-null FWER was 0.0938 at $ETTTT$, whereas replacing the efficacy-null $E$ arm by the truly promising state $A$ produced $ATTTT$ and increased the exact FWER to 0.0956. Here, $E$ denotes the efficacy-null boundary $L_E$, $T$ denotes the toxicity-null boundary $L_T$, and $A$ denotes the maximally favourable alternative. This observed reversal is the practical counterpart of the theoretical distinction above: a complete-null boundary configuration is least favourable for an uncoupled rule, whereas active-count coupling can make a mixed null-and-alternative configuration more adverse.

\begin{table}[!htbp]
\centering
\begin{threeparttable}
\caption{Exact strong familywise-error verification for the three-, four-, and five-dose analyses.}
\label{tab:exact-fwer-main}
\small
\begin{tabular}{rllll}
\toprule
$K$ & Procedure & Complete-null max./worst & Mixed max./worst & Overall FWER max./worst \\
\midrule
3 & Uncoupled & 0.0948 / $EEE$ & 0.0643 / $AEE$ & 0.0948 / $EEE$ \\
3 & AC-coupled & 0.0934 / $TTT$ & 0.0841 / $AEE$ & 0.0934 / $TTT$ \\
\addlinespace
4 & Uncoupled & 0.0942 / $TTTT$ & 0.0715 / $ATTT$ & 0.0942 / $TTTT$ \\
4 & AC-coupled & 0.0969 / $TTTT$ & 0.0921 / $AATT$ & 0.0969 / $TTTT$ \\
\addlinespace
5 & Uncoupled & 0.0889 / $EEEEE$ & 0.0718 / $AEEEE$ & 0.0889 / $EEEEE$ \\
5 & AC-coupled & 0.0938 / $ETTTT$ & 0.0956 / $ATTTT$ & 0.0956 / $ATTTT$ \\
\bottomrule
\end{tabular}%
\begin{tablenotes}[flushleft]
\footnotesize
\item[] \textit{Note:} Overall FWER max. is the maximum FWER over the complete finite boundary set. For each maximum, ``worst'' denotes the boundary configuration at which it is attained.
\end{tablenotes}
\end{threeparttable}
\end{table}

The independent Monte Carlo audits were concordant with the exact calculations. For $K=3$, the maximum simultaneous upper bounds were 0.0971 for Uncoupled and 0.0954 for AC-coupled; for $K=4$ they were 0.0961 and 0.0988, and for $K=5$ they were 0.0899 and 0.0979, respectively. All remained below 0.10. Because the attained exact maxima differ within each $K$, the operating-characteristic comparisons below should be interpreted as comparisons of separately optimized procedures under the same prespecified $\leq0.10$ constraint, not as causal decompositions at exactly matched attained FWER.

\subsection{Operating-characteristic effect of active-count adaptation}
\label{sec:oc-results}

We summarize complementary aspects of retaining truly promising doses. Disjunctive power is the probability that at least one truly promising dose is retained, promising-dose retention is the average marginal retention probability across truly promising doses, conjunctive retention is the probability that all truly promising doses are retained, and exact recovery is the probability that the final retained set equals the true promising set.

Table~\ref{tab:representative-oc-main} summarizes all prespecified $Q_q$ scenarios for the $K=3$, $K=4$, and $K=5$ analyses. Because every truly promising dose has the same data-generating distribution across all three values of $K$, the rows provide a common numerical framework for examining how the contrast between Uncoupled and AC-coupled changes with the number of promising doses. The procedures were nevertheless recalibrated separately for each $K$, so cross-$K$ differences reflect the optimized design at that $K$ rather than the effect of $K$ in isolation.

For $K=3$, the AC-coupled minus Uncoupled retention differences for $q=1,2,3$ promising doses were 0.001, 0.010, and 0.072, and the corresponding conjunctive-retention differences were 0.001, 0.017, and 0.164. Expected-total-sample-size differences were 8.8, 5.2, and 0.9 patients. For $K=4$, the retention differences for $q=1,2,3,4$ were 0.001, 0.006, 0.052, and 0.082, the conjunctive-retention differences were 0.001, 0.010, 0.127, and 0.202, and expected-total-sample-size differences were 1.8, 1.6, 2.3, and 2.4 patients. For $K=5$, the retention differences for $q=1,2,3,4,5$ were 0.013, 0.041, 0.104, 0.119, and 0.177, the conjunctive-retention differences were 0.013, 0.064, 0.190, 0.221, and 0.384, and the corresponding expected-total-sample-size differences were 47.9, 38.5, 28.6, 18.3, and 7.8 patients. Across the three values of $K$, the numerical advantage of active-count adaptation was concentrated in preserving several promising doses simultaneously. Gains were small when only one dose was promising, and the expected-sample-size trade-off was most pronounced in the five-dose, one-promising-dose setting.

\begin{table}[!htbp]
\centering
\begin{threeparttable}
\caption{Operating characteristics under the common $Q_q$ alternatives.}
\label{tab:representative-oc-main}
\footnotesize
\setlength{\tabcolsep}{1.5pt}
\begin{tabular*}{\textwidth}{@{\extracolsep{\fill}}rllrrrrrr@{}}
\toprule
$K$ & Scenario & Procedure & False retention & Disjunctive & Retention & Conjunctive & Exact recovery & Expected $N$ \\
\midrule
3 & 1 promising ($Q_1$) & Uncoupled & 0.001 & 0.813 & 0.813 & 0.813 & 0.812 & 114.8 \\
3 &  & AC-coupled & 0.002 & 0.815 & 0.815 & 0.815 & 0.813 & 123.6 \\
3 & 2 promising ($Q_2$) & Uncoupled & 0.001 & 0.965 & 0.816 & 0.666 & 0.666 & 146.1 \\
3 &  & AC-coupled & 0.003 & 0.968 & 0.826 & 0.684 & 0.681 & 151.3 \\
3 & 3 promising ($Q_3$) & Uncoupled & 0.000 & 0.994 & 0.815 & 0.542 & 0.542 & 177.4 \\
3 &  & AC-coupled & 0.000 & 0.998 & 0.888 & 0.707 & 0.707 & 178.4 \\
\addlinespace
4 & 1 promising ($Q_1$) & Uncoupled & 0.001 & 0.764 & 0.764 & 0.764 & 0.763 & 120.5 \\
4 &  & AC-coupled & 0.001 & 0.765 & 0.765 & 0.765 & 0.764 & 122.2 \\
4 & 2 promising ($Q_2$) & Uncoupled & 0.001 & 0.946 & 0.766 & 0.586 & 0.585 & 153.8 \\
4 &  & AC-coupled & 0.002 & 0.948 & 0.772 & 0.596 & 0.594 & 155.5 \\
4 & 3 promising ($Q_3$) & Uncoupled & 0.000 & 0.987 & 0.765 & 0.448 & 0.448 & 186.8 \\
4 &  & AC-coupled & 0.003 & 0.988 & 0.817 & 0.576 & 0.574 & 189.1 \\
4 & 4 promising ($Q_4$) & Uncoupled & 0.000 & 0.997 & 0.764 & 0.341 & 0.341 & 219.8 \\
4 &  & AC-coupled & 0.000 & 0.998 & 0.846 & 0.543 & 0.543 & 222.3 \\
\addlinespace
5 & 1 promising ($Q_1$) & Uncoupled & 0.001 & 0.708 & 0.708 & 0.708 & 0.708 & 158.3 \\
5 &  & AC-coupled & 0.001 & 0.722 & 0.722 & 0.722 & 0.721 & 206.1 \\
5 & 2 promising ($Q_2$) & Uncoupled & 0.001 & 0.914 & 0.706 & 0.498 & 0.498 & 190.9 \\
5 &  & AC-coupled & 0.002 & 0.933 & 0.747 & 0.562 & 0.561 & 229.4 \\
5 & 3 promising ($Q_3$) & Uncoupled & 0.001 & 0.976 & 0.710 & 0.358 & 0.357 & 223.9 \\
5 &  & AC-coupled & 0.002 & 0.992 & 0.814 & 0.548 & 0.547 & 252.5 \\
5 & 4 promising ($Q_4$) & Uncoupled & 0.000 & 0.993 & 0.708 & 0.251 & 0.251 & 256.7 \\
5 &  & AC-coupled & 0.003 & 0.999 & 0.827 & 0.472 & 0.470 & 275.1 \\
5 & 5 promising ($Q_5$) & Uncoupled & 0.000 & 0.998 & 0.709 & 0.180 & 0.180 & 289.7 \\
5 &  & AC-coupled & 0.000 & 1.000 & 0.886 & 0.564 & 0.564 & 297.4 \\
\bottomrule
\end{tabular*}
\begin{tablenotes}[flushleft]
\footnotesize
\item[] \textit{Note:} False retention is the probability of retaining at least one truly inadmissible dose.
\end{tablenotes}
\end{threeparttable}
\end{table}

The additional sensitivity analyses reinforced the same mechanism-specific interpretation. Across 63 scenarios, the AC-coupled procedure had higher promising-dose retention and higher conjunctive retention in 60; all 48 scenarios containing more than one promising dose favoured AC-coupled on both measures. All 18 association scenarios and all 18 heterogeneous-profile scenarios had positive retention and conjunctive-retention contrasts. The only negative point estimates occurred in three single-promising-dose margin perturbations: $K=3$ at $(p_E,p_T)=(0.45,0.20)$ (retention difference $-0.021$), $K=3$ at $(0.50,0.25)$ ($-0.002$), and $K=5$ at $(0.45,0.20)$ ($-0.014$). Expected total sample size was higher for AC-coupled in all 63 sensitivity scenarios. Thus the additional simulations support the intended role of active-count coupling as a mechanism for preserving multiple viable arms, while showing that a single viable arm need not benefit from that coupling. The Supplementary Material reports the complete sensitivity results and the same $Q_q$ operating characteristics with greater numerical precision.

\section{ABROAD-informed prospective illustration}
\label{sec:abroad-case}

The ABROAD randomized phase II trial provides a clinically relevant three-dose setting for benefit--risk evaluation \citep{tsurutani2021randomized}. Patients with HER2-negative metastatic breast cancer were randomized to nab-paclitaxel 180, 220, or 260~mg/m$^2$ every three weeks. The study planned 42 patients per group (126 in total), and 141 patients were ultimately randomized: 48 to 180~mg/m$^2$, 46 to 220~mg/m$^2$, and 47 to 260~mg/m$^2$. ABROAD itself was not a binary ORR--toxicity screening design: its primary efficacy endpoint and first selection step were based on progression-free survival, followed by a model-based neurotoxicity criterion. The present illustration therefore does not attempt to reproduce the original selection procedure. Instead, it uses the published arm-specific overall response rate (ORR) and observed grade 3/4 sensory-neuropathy rates as clinically grounded margins for a hypothetical prospective binary-endpoint design.

\begin{table}[!htbp]
\centering
\caption{Published ABROAD aggregate margins used as data-generating probabilities in the prospective illustration.}
\label{tab:abroad-main}
\small
\begin{tabular}{lrr}
\toprule
Dose (mg/m$^2$) & ORR (\%) & Observed G3/4 sensory neuropathy (\%) \\
\midrule
180 & 37.8 & 8.3 \\
220 & 44.1 & 8.9 \\
260 & 48.7 & 31.9 \\
\bottomrule
\end{tabular}
\end{table}

We use these margins to define an \emph{ABROAD-informed prospective illustration}, not a retrospective reanalysis of the original trial. The prospective binary endpoints are ORR and grade 3/4 sensory neuropathy. We set $K=3$, a maximum of $N=45$ evaluable patients per active dose, analyses at $n=25,35,45$, and trial-level strong-FWER level $\alpha=0.10$. The formal null boundaries are $\phi_E=\phi_T=0.20$.

A weak Dirichlet working prior with effective sample size one was centered at $(p_E,p_T)=(0.20,0.20)$ under working odds ratio one, giving $(a_{00},a_{01},a_{10},a_{11})=(0.64,0.16,0.16,0.04)$. The monitoring tables were recalibrated using generic $Q_q$ alternatives with promising-dose margins $(p_E,p_T)=(0.40,0.10)$ and odds ratio one; the ABROAD outcome rates were not used to select tuning parameters or integer boundaries. No integer boundary was imposed. After calibration, the frozen tables were verified exactly over the finite boundary set before evaluating the ABROAD-informed scenario.

The induced decision boundaries are shown in Table~\ref{tab:abroad-boundaries-main}. In particular, the Uncoupled rule required at most four toxicities among 45 evaluable patients at the final analysis. AC-coupled used the same terminal toxicity cutoff when one or two doses remained active and allowed at most five toxicities when all three doses remained active. Thus the final observed toxicity cutoffs, 4/45=8.9\% and 5/45=11.1\%, arose from the calibrated posterior rules and were not hard-coded. The corresponding final efficacy cutoffs were 15/45=33.3\% for Uncoupled and for AC-coupled when one or two doses remained active, and 13/45=28.9\% for AC-coupled when all three remained active.

\begin{table}[!htbp]
\centering
\caption{Induced ABROAD-informed decision boundaries under $\phi_E=\phi_T=0.20$. An active dose passes an analysis when both count conditions are satisfied.}
\label{tab:abroad-boundaries-main}
\small
\begin{tabular}{lllc}
\toprule
Analysis $n$ & Procedure & Active doses $m$ & Pass region \\
\midrule
25 & Uncoupled & 1--3 & $x_E\geq6,\;x_T\leq3$ \\
25 & AC-coupled & 1--2 & $x_E\geq6,\;x_T\leq3$ \\
25 & AC-coupled & 3 & $x_E\geq5,\;x_T\leq3$ \\
\addlinespace
35 & Uncoupled & 1--3 & $x_E\geq10,\;x_T\leq3$ \\
35 & AC-coupled & 1 & $x_E\geq9,\;x_T\leq3$ \\
35 & AC-coupled & 2--3 & $x_E\geq9,\;x_T\leq4$ \\
\addlinespace
45 & Uncoupled & 1--3 & $x_E\geq15,\;x_T\leq4$ \\
45 & AC-coupled & 1--2 & $x_E\geq15,\;x_T\leq4$ \\
45 & AC-coupled & 3 & $x_E\geq13,\;x_T\leq5$ \\
\bottomrule
\end{tabular}
\end{table}

For the operating-characteristic scenario, the published margins in Table~\ref{tab:abroad-main} were used only after the monitoring tables had been selected and frozen. Under $\phi_E=\phi_T=0.20$, the 180- and 220-mg/m$^2$ doses are truly promising and the 260-mg/m$^2$ dose is inadmissible because of toxicity. Because the within-patient association between response and sensory neuropathy was not reported, odds ratio one was used for the primary simulation, with 0.5 and 2 examined in sensitivity analyses.

Both frozen tables satisfied exact strong-FWER control. The exact maxima were 8.54\% for Uncoupled and 9.30\% for AC-coupled, both below the prespecified 10\% level. For AC-coupled, the largest mixed-configuration risk was 9.23\% at $AAE$, close to the complete-null maximum, illustrating why mixed configurations must be included in verification even when the overall maximum happens to be complete-null in this example.

Table~\ref{tab:abroad-oc-main} summarizes the primary operating characteristics from 100,000 simulated trials. AC-coupled increased exact recovery of the true retained set $\{180,220\}$ from 24.49\% to 29.88\% and increased the probability of retaining at least one truly promising dose from 74.53\% to 77.83\%. The probability of falsely retaining 260~mg/m$^2$ was 0.011\% for Uncoupled and 0.048\% for AC-coupled. The expected total sample size was 103.5 for Uncoupled and 106.4 for AC-coupled. Both are numerically below the 126 patients originally planned in ABROAD and the 141 patients actually randomized. This comparison is descriptive rather than a direct efficiency comparison because the original trial and the present illustration use different endpoints, monitoring rules, and selection objectives.

\begin{table}[!htbp]
\centering
\caption{ABROAD-informed operating characteristics under the primary within-patient odds ratio of one, based on 100,000 simulations. Probability entries are percentages.}
\label{tab:abroad-oc-main}
\small
\resizebox{\textwidth}{!}{%
\begin{tabular}{lrrrrrrr}
\toprule
Procedure & Exact recovery (\%) & Retain 180 (\%) & Retain 220 (\%) & False retain 260 (\%) & At least one promising (\%) & Expected $N$ & Strong FWER (\%) \\
\midrule
Uncoupled & 24.49 & 46.64 & 52.38 & 0.011 & 74.53 & 103.5 & 8.54 \\
AC-coupled & 29.88 & 50.77 & 56.97 & 0.048 & 77.83 & 106.4 & 9.30 \\
\bottomrule
\end{tabular}%
}
\end{table}

Results of the within-patient association sensitivity analyses are provided in the Supplementary Material.

\section{Discussion}
\label{sec:discussion}

The principal contribution of this work is an exact finite-boundary characterization of strong familywise error for multistage multi-arm efficacy--toxicity screening when armwise decisions are adaptively coupled through the evolving number of active arms. Because the dose-specific null hypothesis is a continuous union region, strong error control must hold over arbitrary mixtures of inadmissible and promising doses rather than only under a complete null. We show that this apparently high-dimensional continuous optimization problem can nevertheless be reduced exactly to a finite set of least-favourable boundary states. For uncoupled armwise rules, the strong FWER has an exact product representation and a complete-null boundary configuration is least favourable. Under active-count coupling, this product structure no longer holds, but the supremum over unrestricted valid arm-specific joint efficacy--toxicity distributions is still attained among configurations in which each arm is represented by an efficacy-null boundary, a toxicity-null boundary, or a maximally favourable alternative. For exchangeable designs, the resulting verification problem further reduces to only $K(K+3)/2$ probability classes. The proposed framework therefore converts a continuous composite-null multiplicity problem into a finite and exactly verifiable one while retaining the adaptive dependence created by active-count monitoring.

The main technical advance is the pathwise argument that establishes this finite least-favourable set despite the dependence induced by active-count coupling. Under the independent-arm sampling model, prespecified per-dose analysis schedules, analysis timing not driven by the efficacy or toxicity outcomes, arm-specific monitoring statistics, and stagewise monotonicity, patient-level couplings can move arbitrary efficacy--toxicity distributions toward favourable boundary states without decreasing the false-retention event. The coupled setting is fundamentally different from the uncoupled case because a favourable change in one arm can keep that arm active and thereby alter the future monitoring boundaries applied to the other arms. This is why the maximally favourable alternative state becomes an essential component of the strong-FWER boundary set. Importantly, the reduction does not require a parametric assumption on the within-patient association between efficacy and toxicity. Once a monitoring table has been selected, its induced integer boundaries are frozen and all required boundary configurations can be evaluated by deterministic finite-state recursion. This separation between design construction and error verification allows simulation to be used flexibly for optimizing operating characteristics while providing an exact strong-FWER guarantee for the final selected design. The framework thereby extends beyond calibration at a limited collection of reference configurations and directly addresses strong error verification for the complete active-count-coupled multistage procedure \citep{mulier2024bayesian,chen2026bop2,yang2024design,tabata2026dose}.

The numerical study demonstrates the practical value of active-count coupling and confirms the theoretical motivation for strong rather than complete-null verification. In the five-dose AC-coupled design, the largest complete-null FWER was 0.0938 at $ETTTT$, whereas the mixed $ATTTT$ configuration had exact FWER 0.0956. Thus, replacing an efficacy-null arm by a truly promising arm increased the trial-level false-retention probability even though the promising arm itself did not contribute to the error event. This directly illustrates why mixed configurations must be included when verifying an active-count-coupled rule. Active-count adaptation was most useful when several promising doses coexisted, where maintaining multiple viable candidates is particularly important for subsequent dose optimization. Because the attained exact maxima differ within each $K$, the operating-characteristic comparisons should be interpreted as comparisons of separately optimized procedures under the same prespecified $\leq0.10$ constraint, not as causal decompositions at exactly matched attained FWER. The same overall pattern across alternative efficacy--toxicity margins, heterogeneous dose profiles, and different within-patient associations further supports the interpretation of active-count adaptation as a mechanism for preserving multiple promising candidates under rigorous error control.

The ABROAD-informed case study provides a complementary assessment of these properties in a clinically meaningful dose-optimization setting. With two promising lower doses and a clearly toxic higher dose, active-count coupling improved preservation of the promising dose set while maintaining a very low probability of retaining the toxic high dose. This is precisely the setting in which preserving more than one viable candidate is useful: rather than prematurely narrowing development to a single dose, the monitoring procedure can retain multiple plausible doses for subsequent benefit--risk evaluation while removing a dose with an unfavourable toxicity profile. The case study also illustrates the practical value of multistage monitoring, because early removal of an unsuitable dose can limit expected enrollment while viable candidates remain under evaluation. The same qualitative conclusion was maintained across the within-patient association sensitivity analyses, indicating that the observed advantage was not driven by the primary working association. Overall, the case study supports the proposed framework as a way to preserve clinically relevant dose options while maintaining rigorous strong-FWER protection.
 
In summary, the proposed framework provides a rigorous connection between adaptive multi-arm efficacy--toxicity monitoring and exact verification of strong FWER. Its central contribution is the characterization of the least-favourable structure for a broad class of outcome- and active-count-monotone procedures, which converts continuous strong-FWER verification into a finite deterministic calculation. The framework clarifies when complete-null verification is sufficient, when mixed null-and-alternative configurations must be considered, and how simulation-based design optimization can be combined with an exact frequentist error guarantee. These guarantees apply to the theorem-compatible class defined by prespecified per-arm analysis schedules, analysis timing not driven by the efficacy or toxicity outcomes, arm-specific monitoring statistics, and stagewise monotonicity; adaptive features that alter these pathwise properties require corresponding verification arguments. By preserving multiple promising candidates while allowing clearly unsuitable doses to be removed during the trial, active-count monitoring is well suited to prospective randomized dose-optimization studies in which benefit--risk assessment may require continued comparison of several viable doses. More broadly, the finite-boundary principle provides a foundation for extending exact error verification to richer adaptive monitoring structures as corresponding pathwise arguments are developed.

\section*{Supplementary material}
A single Supplementary Materials file contains the detailed literature comparison, the detailed finite-state recursion, a state-space scaling summary for the exact verifier, utility-based final selection, the conditional shared-control extension, efficacy-only and categorical-endpoint extensions, complete common-$Q_q$ simulation settings and results for $K=3$, $K=4$, and $K=5$, operating-characteristic sensitivity analyses, exact-verification and independent-validation details, and the ABROAD-informed case-specific calibration and operating characteristics.

\section*{Data and code availability}
No individual patient-level data were used in this methodological study. The ABROAD-informed illustration uses only aggregate quantities reported in the published trial. The R code and reproducibility materials for the numerical studies and the ABROAD-informed illustration are publicly available at GitHub: \url{https://github.com/masahikoji/finite-boundary-strong-fwer}. Version 1.0.0 is permanently archived on Zenodo at \url{https://doi.org/10.5281/zenodo.22121567}. The repository includes the simulation and calibration scripts, configuration files with prespecified scenarios and random-number seeds, deterministic finite-state recursion for exact strong-FWER verification, and scripts for the ABROAD-informed case study.

\bibliographystyle{plainnat}
\bibliography{main}

\end{document}